\documentclass[fleqn,10pt]{wlscirep}
\usepackage[utf8]{inputenc}
\usepackage[T1]{fontenc}
\usepackage{threeparttable}
\usepackage{comment}
\usepackage{siunitx}
\usepackage{url}

\title{Demonstration of nuclear gamma-ray polarimetry based on a multi-layer CdTe Compton Camera}

\author[1,2*]{S. Go}
\author[3,4,2]{Y.~Tsuzuki}
\author[5,2]{H.~Yoneda}
\author[6]{Y.~Ichikawa}
\author[2]{T.~Ikeda}
\author[7]{N.~Imai}
\author[1,2]{K.~Imamura}
\author[2]{M.~Niikura}
\author[8]{D.~Nishimura}
\author[3]{R.~Mizuno}
\author[4]{S.~Takeda}
\author[1,2]{H.~Ueno}
\author[9]{S.~Watanabe}
\author[7,10]{T.~Y.~Saito}
\author[7,2]{S.~Shimoura}
\author[8]{S.~Sugawara}
\author[1,2]{A.~Takamine}
\author[4,3]{T.~Takahashi}
\affil[1]{RIKEN Cluster for Pioneering Research, RIKEN, Saitama, Japan}
\affil[2]{RIKEN Nishina Center for Accelerator-Based Science, RIKEN, Saitama, Japan}
\affil[3]{Department of Physics, The University of Tokyo, Tokyo, Japan}
\affil[4]{Kavli Institute for the Physics and Mathematics of the Universe (WPI), The University of Tokyo, Chiba, Japan}
\affil[5]{Julius-Maximilians-Universit\"{a}t W\"{u}rzburg, Fakult\"{a}t f\"{u}r Physik und Astronomie, Institut f\"{u}r Theoretische Physik und Astrophysik, Lehrstuhl f\"{u}r Astronomie, Emil-Fischer-Str. 31, D-97074 W\"{u}rzburg, Germany}
\affil[6]{Department of Physics, Kyushu University, Fukuoka, Japan}
\affil[7]{Center for Nuclear Study, the University of Tokyo, Saitama, Japan}
\affil[8]{Department of Natural Sciences, Tokyo City University, Tokyo, Japan}
\affil[9]{Institute of Space and Astronautical Science, Japan
Aerospace Exploration Agency, Kanagawa, Japan}
\affil[10]{Atomic, Molecular, and Optical Physics Laboratory, RIKEN, Saitama, Japan}

\affil[*]{go@riken.jp}

\begin{abstract}
To detect and track structural changes in atomic nuclei, 
the systematic study of nuclear levels with firm spin-parity assignments is important. 
While linear polarization measurements have been applied to determine the electromagnetic character of gamma-ray transitions, 
the applicable range is strongly limited due to the low efficiency of the detection system. 
The multi-layer Cadmium-Telluride (CdTe) Compton camera can be a state-of-the-art  
gamma-ray polarimeter for nuclear spectroscopy with the high position sensitivity and the detection efficiency. 
We demonstrated the capability to operate this detector as a reliable gamma-ray polarimeter by using polarized 847-keV gamma rays produced by the $^{56}\rm{Fe}({\it p},{\it p'}\gamma)$ reaction.
By combining the experimental data and simulated calculations, 
the modulation curve for the gamma ray was successfully obtained.
A remarkably high polarization sensitivity  
was achieved, compatible with a reasonable detection efficiency.
Based on the obtained results, a possible future gamma-ray polarimetery is discussed. 

\end{abstract}
\begin{document}
\flushbottom
\maketitle

\section*{Introduction}

In nuclear physics, various exotic phenomena have been found by changing neutron and proton numbers.  
A breakdown of the neutron magic number $N=20$~\cite{Thibault1975, Motobayashi1995} suggested the concept of the ``Island of inversion''.
Shape coexistence in excited states were discussed in the neutron-rich ``doubly-magic'' nuclei $^{78}$Ni~\cite{Taniuchi2019}.
Reflection asymmetry driven by octupole deformation produces parity doublet bands in the odd-mass nuclei~\cite{Dahlinger1988}. 
Sudden shape changes recognized in neutron-rich zirconium isotopes~\cite{Togashi2016,Boulay2020} were interpreted as an occurrence of the quantum phase transition~\cite{Cejnar2010}.
To detect and track these structural changes, 
the systematic study of the energy levels with firm spin-parity assignments is indispensable. 
However, the experimental methods have been strongly limited due to the available statistics of rare isotopes. 
In such circumstances, tentative assignments are often carried out by taking into account the transition strengths derived from experimental data, extrapolation of the systematics from less-exotic nuclei, and theoretical predictions. 
The development of efficient gamma-ray polarimeter, which can be 
applied for the spin-parity assignments in the wide range of nuclei, is expected to open up new perspectives in nuclear physics research.

\begin{figure}[ht]
\centering
\includegraphics[scale=0.60, angle=0]
{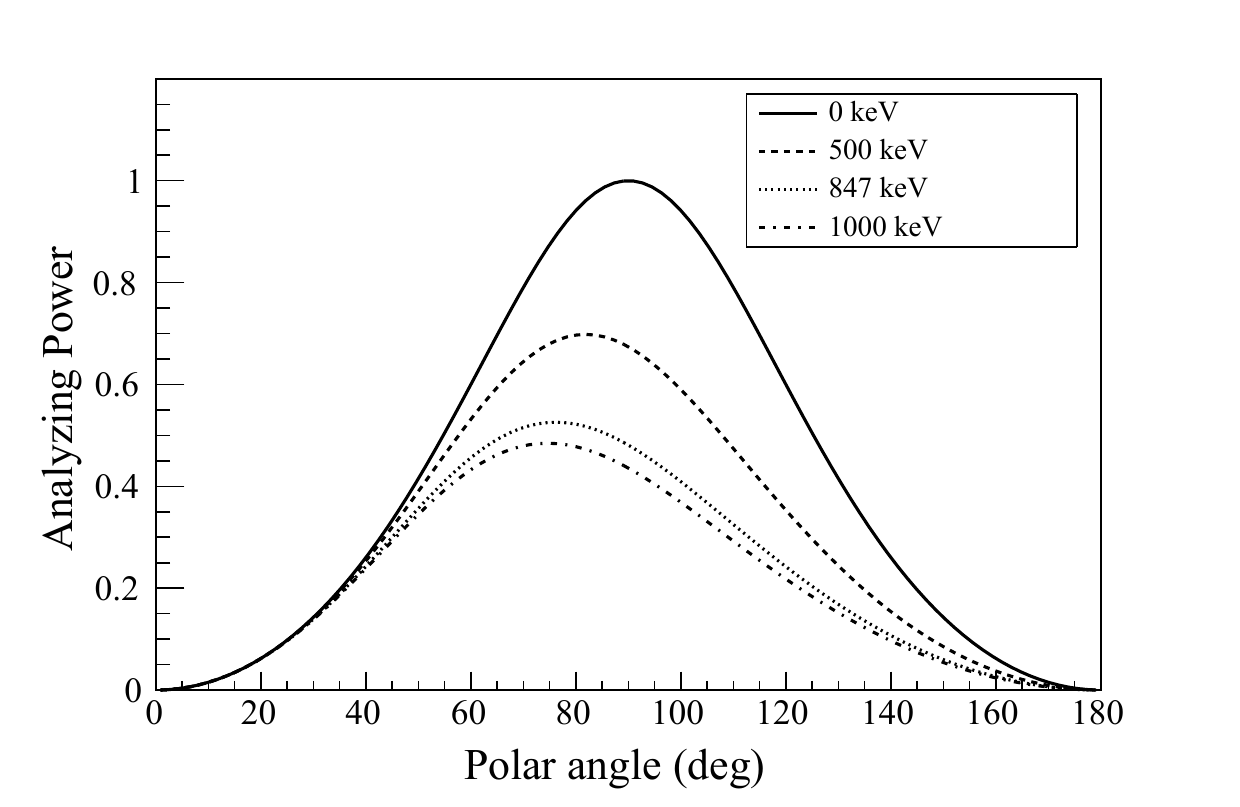}
\caption{Analyzing power depends on the gamma-ray energy.}
\label{fig:analyzingpower}
\end{figure}

Linear-polarization measurements of gamma rays have been applied to determine the electromagnetic character of the nuclear transitions.
The differential cross section for Compton scattering of linear-polarized photons is described by Klein-Nishina formula~\cite{Klein1929}
\begin{equation}
\cfrac{d\sigma}{d\Omega}(\theta, \phi) = \cfrac{r_e^2}{2} \left( \cfrac{E}{E_0} \right)^2 \left(  \cfrac{E}{E_0} + \cfrac{E_0}{E}  - 2\sin^2 \theta \cos^2\phi \right),
\end{equation}
where $r_e$ is the classical electron radius, $E_0/E$ is the ratio of energy of incident and scattered photons, $\theta$ is the polar scattering angle, $\phi$ is the angle between the polarization plane of the incident photon and the Compton scattering plane, respectively. 
The formula represents the anisotropic dependence of the differential cross sections.
While electric dipole ($E1$) radiation is dominant in atomic systems, 
electromagnetic multipole radiation often occurs in nuclear systems due to the actual level structure and the relevant nuclear wave functions. 
Therefore, linear polarization measurements in nuclear spectroscopy play an essential role for spin-parity assignments of the excited states. 
The analyzing power is known to be described as  
\begin{equation}
\Sigma(\theta, E_{\gamma}) = \cfrac{\sin^2 \theta}{\cfrac{E'_{\gamma}}{E_{\gamma}}+\cfrac{E_{\gamma}}{E'_{\gamma}}  -\sin^2 \theta},
\end{equation}
where $E_{\gamma}$ and $E'_{\gamma}$ are energies of incoming and scattered gamma rays. 
The analyzing power as a function of polar scattering angle $\theta$ is shown in Fig.~\ref{fig:analyzingpower}. 
For sub-MeV gamma-ray polarimetry, detection areas which cover around $70^\circ$ are important to achieve high analyzing power.

To obtain the angular distribution of the Compton scattering events, the detection system typically consists of two types of detectors,
the scatterer and the absorber.
In the early stage of the developments, the polarization sensitivity was achieved by combining scintillation counters~\cite{Metzger1950}. 

The high-energy resolution spectroscopy was achieved with different configurations of Ge(Li) spectrometers, such as in the separate crystals~\cite{Broude1969,Bass1972,Butler1973,Ashibe1975,Bass1979,Werth1995}, planer-type crystal~\cite{Filevich1977}, and single crystal
with electrically divided regions~\cite{Werth1995, Hardy1971,Aoki1975,Simpson1983,Schlitt1994,Schmid1998}.
Among those intensive research activities, the advent of the Clover Ge detector~\cite{Duchene1999}, 
which contains four large-volume hyper pure Ge crystals, 
provided the reasonable polarization sensitivities and detection efficiencies~\cite{Jones1995,Kojima2021}.
By selecting separately the scattered events with 
the different crystals,
the detector provided a simple identification of the transition character. 
Recently, gamma-ray tracking arrays such as the Advanced GAmma Tracking Array (AGATA)~\cite{Akkoyun2012} and the Gamma-Ray Energy Tracking In-beam Nuclear Array (GRETINA)~\cite{Paschalis2013} have begun to provide a detailed scattered angular distribution of polarized gamma rays~\cite{Alikhani2012,Bizzeti2015,Morse2022}.
These detector arrays have a few-millimeter position resolution through pulse-shape analysis and signal decomposition. The high-position granularity has significantly improved the performance as a polarimeter without losing the absolute efficiency and the polarization sensitivity even when the detector is placed close enough to the radiation source. 

\begin{figure}[ht]
\centering
\includegraphics[scale=0.18, angle=0]{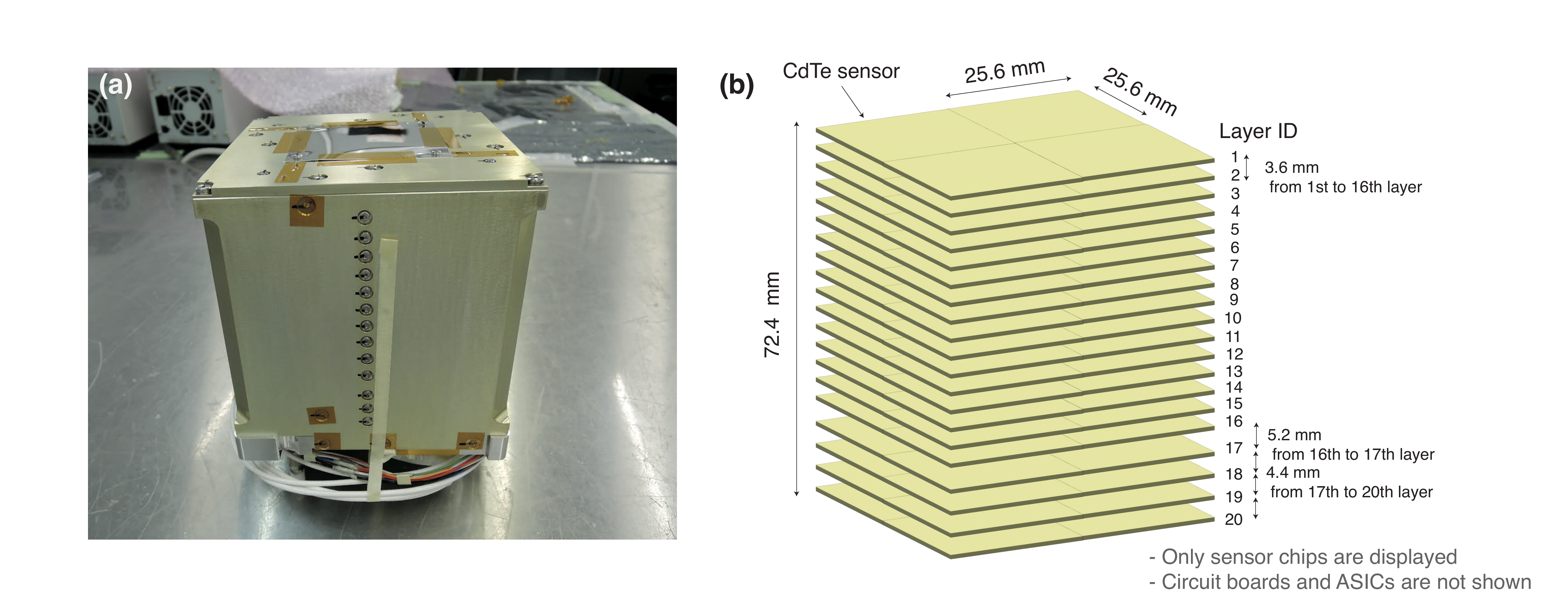}
\caption{(a) Exterior of the multi-layer CdTe Compton camera. (b) Schematic drawing of the inside of the detector. Only sensor chips are displayed, and circuit boards and ASICs are not shown. }
\label{fig:tempcdte20cc}
\end{figure}

Under these circumstances, CdTe-based detectors started showing its capability for Compton polarimetry.
A fine-pitch CdTe-based polarimeter~\cite{Antier2015} showed a high sensitivity for 70-300 keV energy range.
A two-layer configuration based on CdTe pixelized detector~\cite{Moita2019, Moita2021}
indicated fine polarimetric potential of multi-layer CdTe detectors.
Therefore,
multi-layer position sensitive CdTe Compton cameras~\cite{Takahashi2003}
serves as a new device for the linear polarization measurements in nuclear gamma-ray polarimetry.
The detector is expected to offer the high-position resolution, high detection efficiency, and reasonable energy resolution for sub-MeV gamma rays. 
In this paper, we demonstrated the capability to operate this detector as a reliable gamma-ray polarimer by using polarized 847-keV gamma rays produced by proton inelastic reactions.
By combining the experimental data and simulated calculations, a modulation curve 
was successfully obtained. A remarkably high polarization sensitivity with a reasonable detection efficiency is reported. 
Based on the obtained performance as a polarimeter, a future-possible gamma-ray polarimetry is discussed.

\section*{Methods}

\subsection*{Multi-layer position-sensitive CdTe Compton camera}
The multi-layer position-sensitive semiconductor Compton cameras have been developed originally for astronomical observations on board the {\it Hitomi} satellite~\cite{Takahashi2003, Takahashi2018}.
Detectors which employ both Si and CdTe layers~\cite{Watanabe2014,Katsuta2016} 
have brought the technological advances to various studies
such as in the X-ray polarimetry from highly-charged ions on the ground~\cite{Tsuzuki2021,Nakamura2023}, 
the localization of radioactive materials dispersed following the nuclear power plant accident~\cite{Takahashi2012,Takeda2012}, 
the multi-probe tracker in nuclear medicine and small animal imaging~\cite{Takeda2012_2}.  

As a gamma-ray polarimeter, the stacked CdTe layers are suitable~\cite{Takahashi2003,Takahashi2004, Watanabe2003} to achieve high detection efficiency for sub-MeV energy region.
To demonstrate the capability of the gamma-ray polarimetry, a detector which is composed of twenty CdTe layers was employed for the present work.
The detector and the schematic view of the arrangement of the sensors are shown in Fig.~\ref{fig:tempcdte20cc}. 
The CdTe pad devices were produced by ACRORAD Co. Ltd. 
Each sensor has $8\times8$ pixels with the effective area of $25.6\times25.6$~mm$^2$.
Thus, the uniform position resolution of 3.2~mm was achieved over the effective area. 
The CdTe sensors were arranged in a $2\times2$ array for each layer.  
The thickness of each layer is 0.75~mm, and accordingly the total thickness amounts to 15.0~mm.
The distance of each layer is 3.6~mm from 1st to 16th layers, 5.2~mm from 16th to 17th layers, 4.4~mm from 17th to 20th layers, respectively.  
The sensor configuration is similar to that described with more detail in reference~\cite{Watanabe2014}.
A high voltage around 1000~V was applied for the operation.  
The Schottky-barrier diode type, which employed an aluminium (Al) anode and platinum (Pt) cathode, enabled us to apply the high-bias voltage with low-leakage current. 
The temperature around $-18 ^\circ{\rm C}$ were kept with a refrigeration system to 
prevent the sensors from 
%the so-called polarization phenomenon~\cite{Takahashi2001} and also to reduce the leakage current for better noise performance. 
the time instability under bias voltage (known as polarization phenomenon~\cite{Takahashi2001}) 
and also to reduce the leakage current for better noise performance. 

The charge signals from the electrode are amplified and converted to digital signals in the ASICs developed for ${\it Hitomi}$ SGD~\cite{Tajima2010}.
Each channel has two charge sensitive amplifiers. 
One has a short shaping time ($\sim 1 \ \mu \rm s$) for generating a trigger signal, and the other has a longer shaping time ($\sim 3 \ \mu \rm s$) for holding the pulse height. 
The conversion time of the Wilkinson-type analogue-to-digital converters (ADCs) in the ASIC is less than $100~\mu$s. 
The data were collected under the self triggering in zero-suppression mode.
To subtract the common-mode noise level of each event, an ADC of the 32nd pulse height value (half the number of ASIC channels) was detected in every readout timing~\cite{Tajima2010}.

\subsection*{Production and Detection of polarized gamma rays}

\begin{figure}[ht]
\centering
\includegraphics[scale=0.30, angle=0]{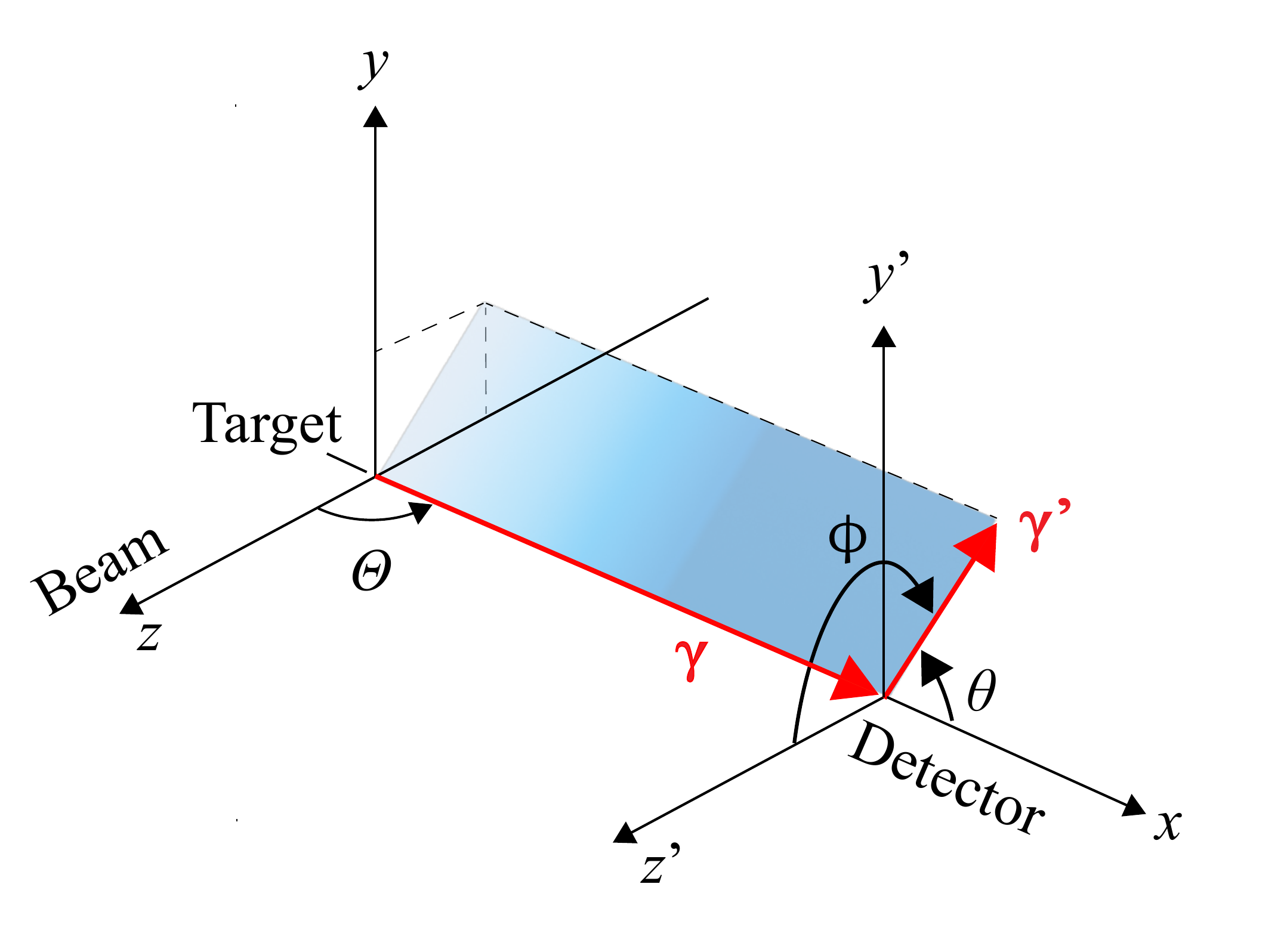}
\caption{Schematic drawing of the measurement of polarized $\gamma$ ray and definition of angles. The target was placed at the origin. $\Theta$ is the polar angle to the beam axis. $(\theta,\phi)$ are polar and azimuthal angles between the incoming and scattered photon directions. }
\label{fig:angle}
\end{figure}

The inelastic reaction, $^{56}\rm{Fe}({\it p},{\it p'}\gamma)$, was selected to 
produce the polarized gamma rays.
The reaction has been known to produce 847-keV gamma ray with nearly 50\% polarization~\cite{Butler1973,Jones1995}.  
The schematic drawing of the measurement and the definition of angles are shown in Fig.~\ref{fig:angle}.
The target was placed at the origin. $\Theta$ is the polar angle to the beam axis. $\theta$ and $\phi$ are polar and azimuthal scattering angles.
The natural iron foil of 10~${\rm \mu}$m was irradiated with the 3.0-MeV proton beam provided by the RIKEN Pelletron accelerator~\cite{Ikeda2022}.
The detector was placed perpendicular to the beam axis, where the high degree of polarization is expected. 
The distance between the target and the front side of the first layer was set to 18.0~cm.
The polar angle corresponded to $\Theta=90^\circ \pm 8^\circ$.
The beam intensity was appropriately tuned during the measurements 
to keep the total count rate around 1~kcps. 
In order to obtain the reference data, a segmented Ge detector (CNS-GRAPE~\cite{Shimoura2004}) was placed on the opposite side of the Compton camera ($\Theta=-90^\circ$).
The intensity of the gamma rays and the experimental asymmetry was monitored in parallel.  

\section*{Results and discussion}

\subsection*{Energy spectra}

\begin{figure}[ht]
\centering
\includegraphics[scale=0.88, angle=0]{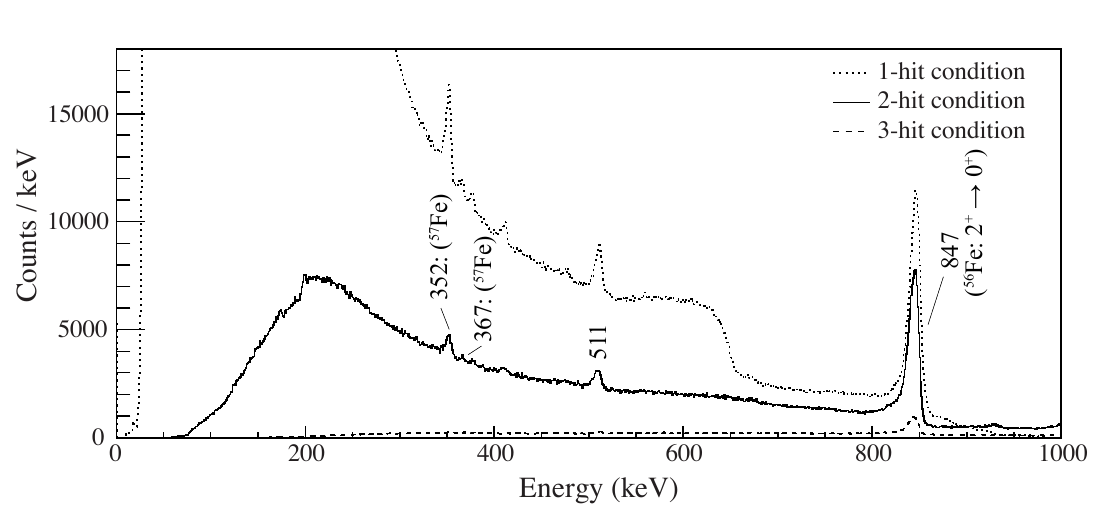}
\caption{Energy spectra with different multiplicity by the multi-layer CdTe Compton camera.The typical count rate of the 847-keV transition was around 3~cps under the 2-hit condition. The data acquisition time was approximately twelve hours. }
\label{fig:espe}
\end{figure}

The calibration parameters from each ADC value to energy of the gamma ray  
were determined from the measurements using standard gamma-ray sources, 
$^{137}$Cs (662~keV), $^{133}$Ba (81, 276, 303, 356~keV), $^{57}$Co (112, 136~keV), $^{22}$Na (511~keV), $^{54}$Mn (835~keV) and $^{154}$Eu (39.5+40.1, 779~keV).
The systematic uncertainties of the calibration based on the sources were estimated to be 0.1$\%$
up to 900~keV.
The energy spectra with different multiplicity are shown in Fig.~\ref{fig:espe}. 
The transition from the first $2^+$ to the $0^+$ ground state in $^{56}$Fe is clearly identified at 847(1)~keV. 
The energy resolution was 1.3\% (FWHM) for the transition. 
There was no significant degradation on the energy resolution depends on the multiplicity.  
Low-intensity peaks at 352 and 367~keV were both assigned as the transitions from the excited state at 367~keV in $^{57}$Fe, 
which is inherent in the iron target with the natural abundance of 2.1\%. 
The average saturation energy of each pixel was estimated to be 890~keV, 
and about 10\% of total number of pixels could not measure the 847-keV transition in a single interaction.
However, the effect was negligible in Compton scattering events, 
which split the total energy in several pixels. 
The pixels which had the readout issue were also taken into consideration both in the data analysis and the simulations to compare them in a consistent way.  
The relative efficiencies from a single-hit event to 2-hit and 3-hit events to produce the full energy peak at 847~keV were estimated to be 63.8(3)$\%$ and 7.6(1)$\%$, respectively. 
The 2-hit events which produced the 847-keV peak were used for the analysis.

\subsection*{Azimuthal distributions and degree of polarization}

The degree of polarization of incoming gamma rays is obtained by comparing the azimuthal angular distributions of the measurement and the simulated 
calculation. 
The optimal azimuthal angle which satisfies the kinematics of Compton scattering is selected in the 2-hit event. 
In the procedure, the detector responses are normalized by the aid of the energy spectra in the early stage, and then the degree of polarization is estimated by the maximum-likelihood method. 
The modulation curve is obtained by taking the ratio of the measured and the simulated azimuthal angular distribution.
The amplitude of the modulation curve indicates the polarimetric performance of the detector.

We utilized the {\ttfamily ComptonSoft} toolkit~\cite{Odaka2010}, which employed {\ttfamily Geant4} simulation framework~\cite{Agostinelli2003} for physical calculations.  
More details on the method can be found in a previous work on hard X-ray scattering polarimetry~\cite{Tsuzuki2021}.

First, we defined the divisions to minimize the effect of the noisy sensor regions and the background.
After defining the division, we constructed $\phi$ histograms designated as $d_{ij}$,
where $i$ denotes the bin number of the $\phi$ histogram and $j$ denotes the index of division. 
The normalization factors $\nu_j$ for each division were obtained by fitting 
the measured energy histograms ($e_{j}^{\rm (exp)}$) with the simulated histogram
obtained by
\begin{equation}
e^{\rm (model)}(P) = \nu_{j} [ (1-P)e^{(0)} + Pe^{(100)}] +  e^{\rm (bg)}, 
\end{equation}
where $P$ is the degree of polarization, $e^{(0)}$ and $e^{(100)}$ represent the simulated energy spectra with unpolarized, 
100$\%$-linearly polarized gamma rays 
and $e^{\rm (bg)}$ represents the background energy histogram from the measurements.
The background spectra were appropriately scaled by the live time of the measurements. 
For the iteration purpose, the degree of polarization was set to an initial value.

Second, the logarithmic likelihood $M(P)$ was computed with the fixed $\nu_j$ given by 
\if0
\begin{equation}
M(P) = \Sigma_{ij} (d^{\rm (exp)}_{ij} \ln d^{\rm (model)}_{ij} - d^{\rm (model)}_{ij}) + \rm const.,
\end{equation}
\fi

\begin{equation}
M(P) = \Sigma_{ij} \ln p_{ij} (d_{ij}^{\rm (exp)};d_{ij}^{\rm (model)}) = \Sigma_{ij} [ d_{ij}^{\rm (exp)} \ln d_{ij}^{\rm (model)}-d_{ij}^{\rm (model)}-\ln(d_{ij}^{\rm (exp)})!]
\end{equation}

where 
\begin{equation}
d^{\rm (model)}_{ij}(P) = \nu_{j} [ (1-P)d^{(0)}_{ij} + Pd^{(100)}_{ij}] +  d^{\rm (bg)}_{ij}.
\end{equation}
For here, the Poisson distribution ($p_{ij}$) is assumed for $d^{\rm (exp)}_{ij}$ with mean of $d^{\rm (model)}_{ij}$.
The procedure was iterated by updating the degree of polarization $P$ until $M(P)$ reached its maximum.
In addition to the iteration process, the optimal polar scattering angle was investigated to increase the polarimeter performance. 
As a result, we found out that the restriction of the polar scattering angle ($40^\circ \leq \theta \leq 120^\circ$) significantly improved the polarization sensitivity with a certain loss of the detection efficiency.

\begin{figure}[ht]
\centering
\includegraphics[scale=0.80, angle=0]{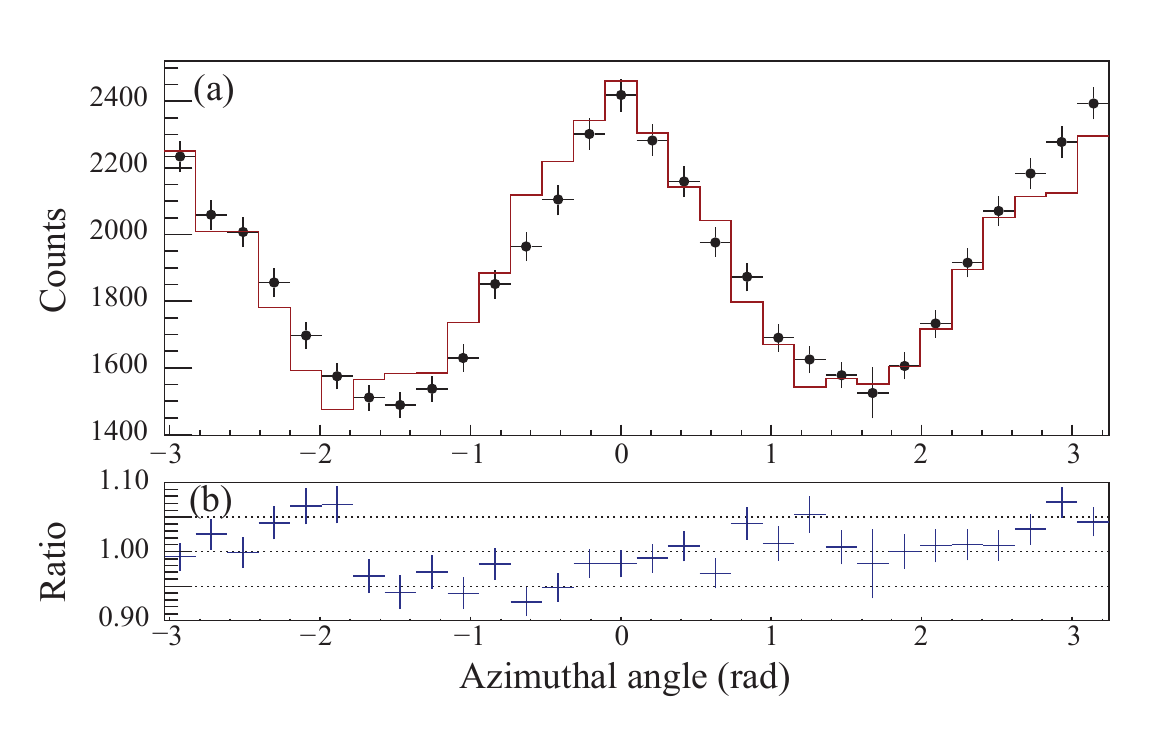}
\caption{(a) Measured azimuthal angular distributions for the 847-keV gamma rays. The black dots and red line represent experimental data and the simulated distributions, respectively. 
The normalization factors were applied into the simulation data before the cut of the polar scattering angle. 
The polar scattering angle was restricted to be $40^\circ \leq \theta \leq 120^\circ $ both for experimental and simulated distributions. (b) Ratio between the experimental data and the simulated results. 
}
\label{fig:hist-phi}
\end{figure}

The measured azimuthal angular distribution and the best-fit simulated results are shown in Fig.~\ref{fig:hist-phi}.
The figure shows the source (black dots) and the simulated (red line) distributions and the ratio between experimental and simulated values. 
As a result, the degree of polarization was determined to be $P=0.57(4)$.
The positive sign of the degree of polarization is defined as an electric transition character, and the known electric character was reproduced~\cite{Butler1973, Jones1995}.  
The uncertainty of the degree of polarization was dominated by systematic one associated with different event selections. 
The systematic uncertainty originating from the detector rotation was also estimated by our simulation, and was negligibly small compared to the precision in the present work.
A strength of this method is the deduction of the degree of polarization without obtaining the angular-intensity distribution with different polar angles~\cite{Bass1979}.

\subsection*{Modulation curve and Polarization sensitivity}

\begin{figure}[h]
\centering
\includegraphics[scale=0.60, angle=0]
{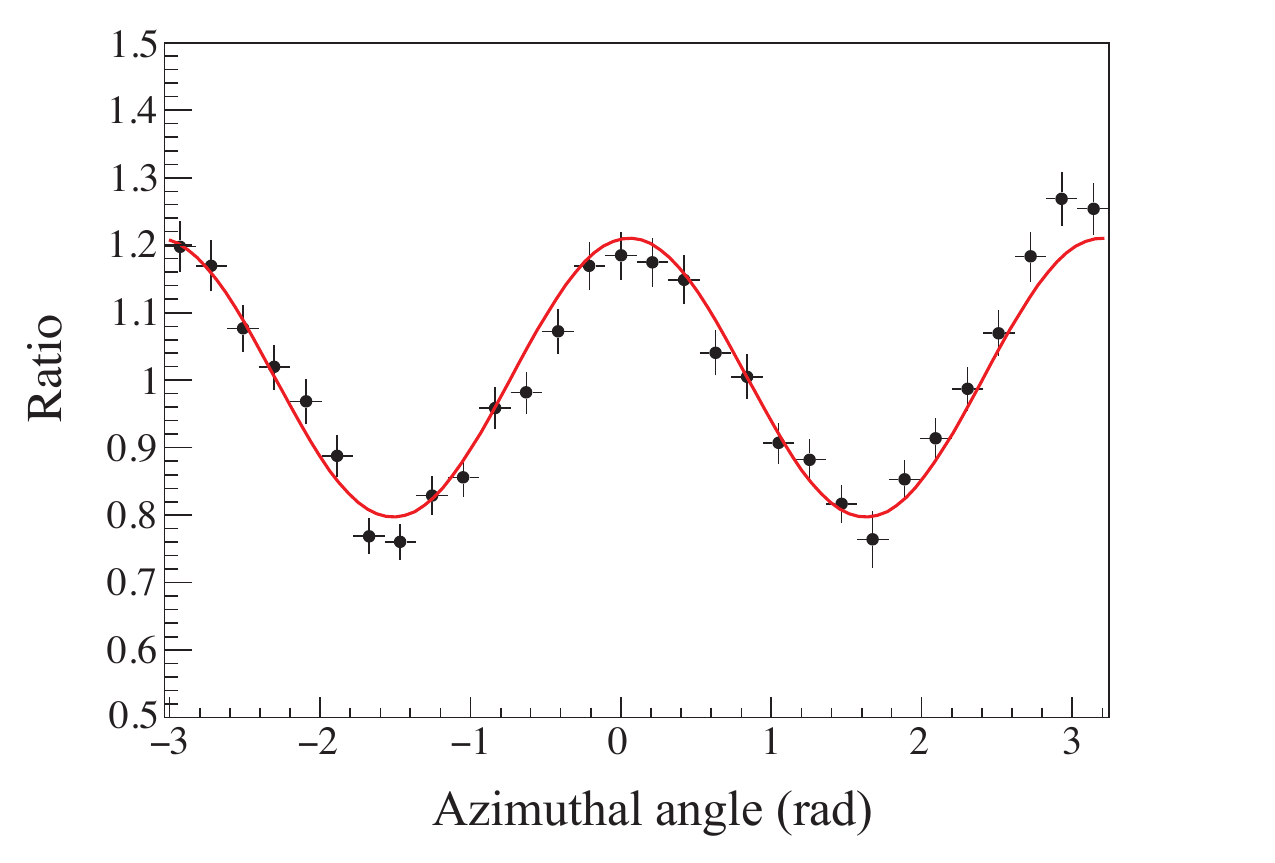}
\caption{Modulation curve for the 847-keV gamma ray. The black dots and red line represent experimental results and the fitted line with $f(\phi) = A (1 + Q' \cos(2(\phi - \phi_0)) )$. The obtained parameters are $A = 1.004(6)$, $Q' = 0.203(8)$, and $\phi_0 = 0.06(3)$, respectively.
}
\label{fig:curve-cos}
\end{figure}

The modulation curve for the 847-keV gamma ray is shown in Fig.~\ref{fig:curve-cos}. 
The curve was obtained by taking the ratio of the measured- and the simulated-azimuthal angular distribution with the obtained degree of polarization, as expressed:
\begin{equation}
f_{\rm modulation}(\phi) = \frac{d^{\rm (exp)}(\phi)-d^{\rm (bg)}(\phi)}{d^{\rm (0)}(\phi)}.
\end{equation}
The curve was fitted with the function $A(1+Q'\cos2(\phi-\phi_0))$
where $A$, $Q'$, and $\phi_0$ are the fitting parameters.
$Q'$ is called modulation factor, which is proportional to the degree of polarization.  
The modulation factor indicates the maximum experimental asymmetry observed in the modulation curve with a given degree of polarization. 
In the present work, 
we obtained $Q' = 0.203(9)$ with $P=0.57(4)$ from the fitting.
The polarization sensitivity for 100\% polarized photon is calculated to be $Q=Q'/P=0.35(4)$ at 847~keV.

To compare the polarization sensitivity with those of other detectors,  
the interpolation has been conducted by scaling the ideal sensitivity with a linear function 
$ Q (E_\gamma) = Q_0(\theta=90^\circ, E_\gamma) \times (aE_{\gamma}+b )$,
where $a$ and $b$ are fitting parameters from experimental data.  
The theoretical limit of the sensitivity ($Q_0$) is calculated by assuming the point-like scatterer and the absorber~\cite{Schlitt1994} as expressed:
\begin{equation}
Q_0 (\theta=90^\circ, E_\gamma) = \frac{1+E_{\gamma} /m_{\rm e}c^2}{1+E_{\gamma} /m_{\rm e}c^2+(E_{\gamma} /m_{\rm e}c^2)^2}, 
\end{equation}
where $m_e c^2$ is electron rest energy. 
For example, the sensitivities of GRETA and Clover detectors can be calculated as $Q(847 \ \rm keV)=0.200(5) \  (a=3.25(7)\times10^{-4}$, $b=0.131(6)$~\cite{Morse2022}) and $0.229 \  (a=1.470\times10^{-4}$, $b=0.3404$~\cite{Kojima2021}), respectively. 
The similar sensitivity, $Q(\rm 847 ~keV)=0.32$, was reported by POLALI~\cite{Werth1995} which adopted a low efficiency germanium crystal as a scatterer.  
Regarding polarimeters adopted CdTe sensors, a modulation factor 0.78 was reported in 200--300~keV region with a single layer, fine-pitch CdTe sensor~\cite{Antier2015}, and 0.13 was reported in 278~keV with two-layer configurations~\cite{Moita2019, Moita2021}, respectively. The estimation of the polarimetric performance of the energy range will be important for quantitative comparison.
While the polarization sensitivity is an important design parameter for polarimeters, the sensitivity does not necessarily take into account the aspect of the detection efficiency. 
One can achieve a higher polarization sensitivity by imposing more strict constraints on the event selection. 
For example, the narrower polar scattering angle 
($60^\circ \leq \theta \leq 100^\circ$) for the present data set  
provides a higher sensitivity of $Q (\rm 847 \ keV)=0.40(4)$, 
but results in the loss of 65\% of the detection efficiency.

In position-insensitive detectors, 
the polarization sensitivity and the coincidence efficiency are typically in a trade-off relationship for a given detector volume and segmentation.
To consider this aspect, a figure of merit~\cite{Logan1974,Simpson1983} has been used as a quality description for polarimeters
$F = \epsilon Q^2$, where $\epsilon$ is the 
detection efficiency of Compton scattering events.
The increase of the figure of merit results in a proportional decrease of the time to achieve required statistical significance.
This efficiency is not necessarily identical with the intrinsic absolute coincidence efficiency, 
and here needs to consider the loss of the statistics by the event selection.
With the use of {\ttfamily ComptonSoft} toolkit~\cite{Odaka2010}, the absolute coincidence detection efficiency for 847-keV gamma-ray in the present setup was estimated to be $\epsilon_{\rm coin} = 3.0\times10^{-5}$ by considering the distance between the target and the detector, obtained the size of the radiation spot, and passive materials around the detector.
For the present work, we imposed the restriction on the polar scattering angle
($40^\circ \leq \theta \leq 120^\circ$) to maximize the figure of merit. 
The polarization sensitivity increased by 1.5 times in exchange for a 36\% the efficiency loss due to event selection.
By considering these facts, 
we adopted $\epsilon \sim 1.9 \times 10^{-5}$ for the efficiency.   
The figure of merit was estimated to be $F\sim 2.4 \times 10^{-6}$.
Note that the efficiency took into consideration the pixels which were not used for 
the analysis, but the materials such as the beam duct and the refrigerator were not considered.
The performance is summarized in Table~\ref{table:performance} with those of other gamma-ray polarimeters.
The highest polarization sensitivity for 847~keV was demonstrated by the multi-layer CdTe detector with the comparable figure of merit. The strength of the position sensitive detector lies also in the capability to adjust its sensitivity for the transition of interest by optimizing event selection.

\subsection*{Possible future gamma-ray polarimetry}

In the present work, a remarkably high polarization sensitivity by the multi-layer CdTe Compton Camera was demonstrated for the 847-keV gamma ray.
In Table~\ref{table:performance}, one notices the significantly high figure of merit by the tracking Ge detector~\cite{Morse2022}. The high value is attributed to the high-detection efficiency of the array ($\epsilon= 5.3(1)\times10^{-2}$). Increasing the absolute detection efficiency will be important for 
reliable gamma-ray polarimetry. 
Several possibilities for possible future gamma-ray polarimetry are discussed. 

\begin{threeparttable}[htb]
 \caption{
 Comparison of the performance of CdTe Compton camera and other gamma-ray polarimeters. 
 }
 \label{table:performance}
 \centering
  \small
  \begin{tabular}{llrcSc}
   \hline
Detector & Reference & {Energy ($E_{\gamma}$)} & Reaction & {Polarization Sensitivity $(Q)$} & Figure of Merit ($F$) \\ 
\hline
CdTe Compton Camera & This work & 847 & $^{56}$Fe$(p,p'\gamma)$ & 0.35 & $2.4\times 10^{-6}$ \\
\hline 
CdTe (Single layer, Fine pitch) & Antier ${\it et \ al}$.~\cite{Antier2015} & 200--300 & Syncrotron radiation & 0.78\tnote{a} & - \\
CdTe (Two layers) & Moita ${\it et \ al}$.~\cite{Moita2019} & 278 & Syncrotron radiation & 0.13\tnote{a} & - \\
\hline
POLALI & Werth ${\it et \ al}$.~\cite{Werth1995}  &  847 & $^{56}{\rm Fe}(p,p'\gamma)$ & 0.32 & $4.0\times 10^{-6}$ \\
MINIPOLA & Werth ${\it et \ al}$.~\cite{Werth1995}  & 847 & $^{56}{\rm Fe}(p,p'\gamma)$ & 0.07 & $2.0\times 10^{-7}$ \\
Gammasphere & Schmid ${\it et \ al}$.~\cite{Schmid1998} &  847  & $^{56}{\rm Fe}(p,p'\gamma)$   & 0.047 &  -  \\
Clover Ge & Kojima ${\it et \ al}$.~\cite{Kojima2021}    & 847 &  - & 0.229\tnote{b} & - \\ 
    GRETINA & Morse ${\it et \ al}$.~\cite{Morse2022}  & 847 &  - & 0.200\tnote{c} & - \\
\hline
DAGATA & Alikhani ${\it et \ al}$.~\cite{Alikhani2012}     & 1173  &  $^{60}$Co  & 0.228 & - \\ 
Clover Ge & Kojima ${\it et \ al}$.~\cite{Kojima2021}    & 1173  &  $^{60}$Co  & 0.181 & $4.7\times 10^{-7}$ \\ 
\hline
DAGATA & Alikhani ${\it et \ al}$.~\cite{Alikhani2012} & 1332 & $^{60}$Co  & 0.192 & - \\
Clover Ge & Kojima ${\it et \ al}$.~\cite{Kojima2021}    & 1332 &  $^{60}$Co  & 0.197 & $5.5\times 10^{-7}$ \\ 
\hline
POLALI & Werth ${\it et \ al}$.~\cite{Werth1995}  & 1368 & $^{24}{\rm Mg}(p,p'\gamma)$ & 0.30 & $1.8\times 10^{-6}$ \\
MINIPOLA &  Werth ${\it et \ al}$.~\cite{Werth1995} & 1368 & $^{24}{\rm Mg}(p,p'\gamma)$ & 0.05 & $2.0\times 10^{-8}$ \\
  Gammasphere & Schmid ${\it et \ al}$.~\cite{Schmid1998} & 1368  & $^{24}{\rm Mg}(p,p'\gamma)$   & 0.043 & $1.7\times 10^{-6}$   \\
    GRETINA & Morse ${\it et \ al}$.~\cite{Morse2022}  & 1368 & $^{24}{\rm Mg}(p,p'\gamma)$ & 0.196 & $2.0\times 10^{-3}$ \\
\hline
  \end{tabular}
  \begin{tablenotes}
  \item[a] estimated as a modulation factor.
  \item[b] interpolated by ($a=3.25\times10^{-4}$, $b=0.131$)~\cite{Morse2022}. 
 \item[c] interpolated by ($a=1.470\times10^{-4}$, $b=0.3404$: setup 
 I\hspace{-1.2pt}I)~\cite{Kojima2021}.
 \vspace{2mm}
  \end{tablenotes}
\end{threeparttable}

First, achieving the large solid angle by modifying the present experimental setup improves the figure of merit. 
In the present work, a commercial refrigeration system to reduce the leakage current was employed. The system prevented the detector from placing close enough to the target.
Minimizing physical conflicts can improve the absolute detection efficiency.
If the distance between the target and the detector is shortened from 18.0~cm (current distance) to a few centimeter, more than 10-times larger solid angle will be achieved.

Second, adding the new CdTe sensors on the side close to the stacked layers will increase the coverage of the scattered gamma ray and analyzing power.  
The technology of stacking layers on the side has already been demonstrated such as in the Si/CdTe Compton camera~\cite{Watanabe2014}. 
In the present study, we found out that the restriction on the polar scattering angle ($40^\circ \leq \theta \leq 120^\circ$) increased the polarization sensitivity and maximized the figure of merit for the transition, 
compatible with a reasonable detection efficiency. 
Adding the side detectors will provide a better coverage for the polar scattering angle.

Lastly, the development of large volume of the CdTe sensors and high-position resolution are quite important for the next-generation gamma-ray polarimetry. 
With recent detector developments, CdTe layers with 
high-position resolution became possible with double-sided strips. 
As an example, the spacial resolution of 250~$\mu$m was recently reported ~\cite{Yabu2021}. 
By improving the position resolution, the close placement to the target will be achieved without the loss of the polarization sensitivity.

For an another perspective, 
the detector with ultrahigh position resolution will simultaneously provide the detailed polar-angle intensity distribution, which has been conventionally measured by changing the angle of detectors or employing the large detector arrays. 
The intensity distribution independently provides the information about the degree of polarization 
at $\Theta=90^\circ$ by fitting it with Legendre polynomials~\cite{Bass1979}. 
The development of high-position resolution detector will serve for a highly-efficient gamma-ray polarimeter not only for rare isotopes,
but also for the systematic study of the degree of polarization, which would 
allow us to gain insight into the population patterns of magnetic substates in excited states.

\subsection*{Summary and Conclusion}

We demonstrated the gamma-ray polarimetry with a multi-layer CdTe detector. 
The polarized 847-keV gamma rays were produced by the $^{56}\rm{Fe}({\it p},{\it p'}\gamma)$ reaction at 3.0-MeV proton beam energy.  
By combining the experimental data and the simulated calculations,
the modulation curve for the transition was successfully obtained. 
The highest polarization sensitivity was achieved for the transition with a reasonable detection efficiency. 
Based on the obtained performance, Compton polarimeter improvements for nuclear physics was discussed.

\section*{Data availability}
The data used and analyzed during the present study available from the corresponding author on reasonable request.

\section*{Acknowledgements}

This work was supported by a Grant-in-Aid for Scientific Research on Innovative Area ``Toward new frontiers: Encounter and synergy of state-of-art astronomical detectors and exotic quantum beams'' and KAKENHI Grant Numbers 18H05463, 20H00153, 22J12155 and 23K03444. 
The experiment was performed at the Pelletron facility (joint-use equipment) at the Wako Campus, RIKEN.
H.Y. acknowledges support by the Bundesministerium f\"{u}r Wirtschaft und Energie via the Deutsches Zentrum f\"{u}r Luft- und Raumfahrt (DLR) under contract number 50 OO 2219.
R.M. was supported by the Forefront Physics and Mathematics Program to Drive Transformation (FoPM), a World-leading Innovative Graduate Study (WINGS) Program, and a JSR Fellowship at the University of Tokyo.

\section*{Author contributions}
S.G. contributed to the conceptualization, methodology, investigation, and wrote the original manuscript. 
Y.T. and H.Y performed the offline data analyses, simulations, and made data interpretation. 
S.G., Y.T., H.Y., T.I., K.I. and A.T. prepared the experimental setup around the target.
Y.T., H.Y., S.T., S.W. and T.T. were responsible for setting up the CdTe Compton camera.  
Y.I., N.I., M.N., D.N, R.M., T.Y.S. S. Shimoura and S.Sugawara were responsible for setting up the CNS-GRAPE. 
T.I. operated the Pelletron accelerator and provided the proton beam. 
H.U. and T.T. supervised the project. 
All authors reviewed the manuscript.

\section*{Competing interests}

The authors declare no competing interests.

\end{document}